\documentclass[runningheads,a4paper]{llncs}
\usepackage{etex}
\reserveinserts{28}
\pdfoutput=1

\usepackage[linesnumbered,lined,boxed,commentsnumbered,ruled]{algorithm2e}

\SetAlFnt{\small}
\SetAlCapFnt{\small}
\SetAlCapNameFnt{\small}
\SetAlCapHSkip{0pt}
\IncMargin{-\parindent}
\usepackage{ textcomp }
\usepackage{ comment }
\usepackage[color=yellow!60,textsize=footnotesize,obeyDraft,draft]{todonotes}
\usepackage{colortbl}
\usepackage{booktabs}
\usepackage{capt-of}
\usepackage{makecell}
\usepackage{textcomp }
\usepackage{multirow}
\usepackage{listings}
\usepackage{graphicx}
\usepackage{colortbl}
\usepackage{booktabs}
\usepackage{amsmath}
\usepackage[font=footnotesize,labelfont=bf]{subcaption}
\usepackage[hidelinks]{hyperref}
\usepackage[capitalize]{cleveref}
\usepackage{tikz}
\usepackage{pgfplots}
\usepackage[para,online,flushleft]{threeparttable}
\usepackage{multirow}
\usepackage{wasysym}
\usepackage{amssymb}
\usepackage{enumitem}
\usetikzlibrary{arrows}
\usepackage{subcaption}
\usepackage{breakcites}
\newcounter{mnotei}
\newcolumntype{L}[1]{>{\raggedright\let\newline\\\arraybackslash\hspace{0pt}}m{#1}}
\newcolumntype{C}[1]{>{\centering\let\newline\\\arraybackslash\hspace{0pt}}m{#1}}
\newcolumntype{R}[1]{>{\raggedleft\let\newline\\\arraybackslash\hspace{0pt}}m{#1}}

\newcommand{\includegraphicsmaybe}[2]{
    \IfFileExists{#2}{\includegraphics[#1]{#2}}{
    \detokenize{File #2 is missing, maybe you need to run plots.py?}
}}

\begin{document}
\bibliographystyle{ieeetr}
\mainmatter

\title{An Agency-Directed Approach to Test Generation for Simulation-based Autonomous Vehicle Verification}

\titlerunning{A Multi-Agent Systems Approach to Test Generation}

\author{Greg Chance\inst{1,3}, Abanoub Ghobrial\inst{1,3}, Severin Lemaignan\inst{2,3}, \\Tony Pipe\inst{2,3}, Kerstin Eder\inst{1,2}}

\authorrunning{G. Chance et al.}
\institute{University of Bristol, Bristol, UK \and University of the West of England, Bristol, UK \and Bristol Robotics Laboratory, University of the West of England, Bristol, UK}
\tocauthor{Authors' Instructions}
\maketitle

\makeatletter
\renewcommand\subsubsection{\@startsection{subsubsection}{3}{\z@}%
                       {-18\p@ \@plus -4\p@ \@minus -4\p@}%
                       {4\p@ \@plus 2\p@ \@minus 2\p@}%
                       {\normalfont\normalsize\bfseries\boldmath
                        \rightskip=\z@ \@plus 8em\pretolerance=10000 }}
\makeatother

\begin{abstract}
Simulation-based verification is beneficial for assessing otherwise dangerous or costly on-road testing of autonomous vehicles (AV). This paper addresses the challenge of efficiently generating effective tests for simulation-based AV verification using software testing agents. The multi-agent system (MAS) programming paradigm offers rational agency, causality and strategic planning between multiple agents. We exploit these aspects for test generation, focusing in particular on the generation of tests that trigger the precondition of an assertion. On the example of a key assertion we show that, by encoding a variety of different behaviours respondent to the agent's perceptions of the test environment, the agency-directed approach generates twice as many effective tests than pseudo-random test generation, while being both efficient and robust. Moreover, agents can be encoded to behave naturally without compromising the effectiveness of test generation. Our results suggest that generating tests using agency-directed testing significantly improves upon random and simultaneously provides more realistic driving scenarios.
\end{abstract}

\section{Introduction} \label{s:introduction}

Verification is the process used to gain confidence in the correctness of a system with respect to its requirements~\cite{bergeron2012writing}. Testing is a technique that can be used to achieve this by showing that the intended and actual behaviours of a system do not differ and detecting failures against the requirements in the process~\cite{utting2012taxonomy}.

Using simulation to test autonomous driving functions in safety critical scenarios benefits from full control over the environment, where road layouts, weather conditions, a variety of road users and other driving scenario parameters can be directed to achieve specific test targets. These tests may aim to provide evidence to regulators of the functional safety of the vehicle or its compliance with commonly agreed upon road conduct, such as the Vienna convention~\cite{ViennaConv}, typically implemented at national level as a set of rules~\cite{codes2015highway}, road traffic laws and penalties~\cite{RoadTraffic1988}.

Verification of complex systems is challenging. In semiconductor design, for example, it has long been recognised that verification can take up to 70\% of the design effort~\cite{arden2002international}, with the largest part still being achieved with simulation-based techniques. The testbench is the code used to drive a stimulus sequence into the Design under Verification (DUV) while observing input protocols. It also records coverage and checks the DUV's response. The testbench provides a completely closed environment from the DUV's perspective. Simulators are used to execute testbenches. Automation plays a critical role in achieving verification targets efficiently and effectively.

Coverage-driven verification is a systematic, goal-directed simulation-based verification method~\cite{HVC2015} that offers a high degree of automation and is capable of exploring systems of realistic detail under a broad range of environment conditions. Because exhaustive simulation is intractabe due to the vast parameter space, the remaining challenge is in strategically selecting the (ideally smallest set of) test cases that result in the highest level of confidence in the design's correctness. Automating test generation has been the focus of research for decades, giving rise to a variety of coverage-directed stimulus generation techniques that exploit formal methods, genetic programming and machine learning~\cite{Ioannides:2012}. 

Compared to semiconductor design verification, AV verification faces even bigger challenges, including automatic test generation. It is well known that few of the valid tests are actually interesting from a verification point of view. Estimates vary, but demonstrating AV safety with a confidence of 95\% that the failure rate is at most 1.09 fatalities per 100 million miles driven would take 275 million miles, equivalent to 12.5 years for a fleet of 100 AVs~\cite{kalra2016driving}. This figure is based on the number of roads fatalities in the US, and would need to be adjusted taking into consideration local statistics on road safety, e.g.\ the number of road fatalities per billion vehicle-km in the UK has been half that of the US in 2018~\cite{ITFroadSafety2018}. Ways must be found to test the scenarios of interest without needing millions of miles of driving or billions of miles of simulated driving~\cite{korosec2019waymo}. In particular, simulation-based testing offers the opportunity to increase the number of otherwise rare events~\cite{Koopman2018} in order to determine whether the AV handles such rare events appropriately. But automation can also apply to not just the process but the method of test generation which is the focus of this paper.

\begin{figure}[!t]
	\centering
\includegraphics[width=0.98\textwidth]{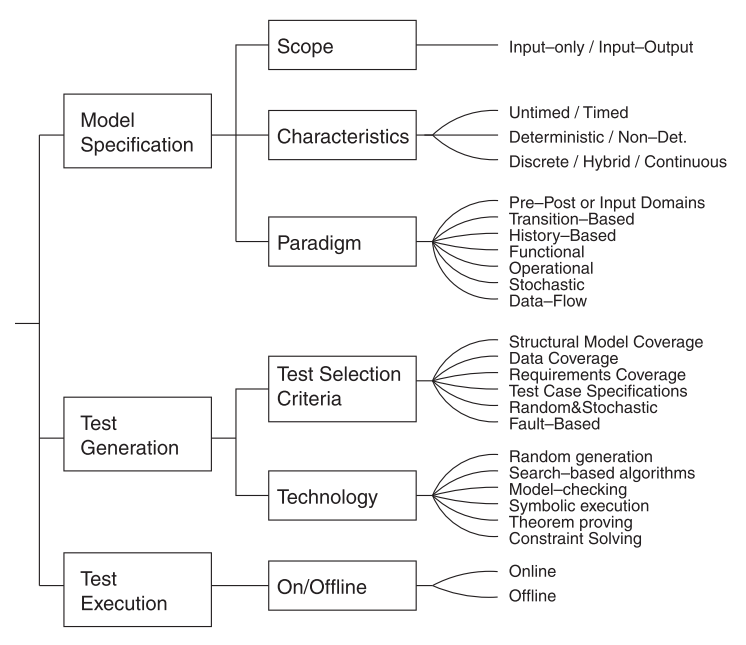}
	\caption{Taxonomy of model-based test generation, from~\cite{utting2012taxonomy}.}
	\label{f:taxonomy}
\end{figure}

Considering the AV as a DUV, the challenge is to generate tests that interact with the AV over a period of time, thereby creating an environment in which the AV needs to respond to the received stimulus while making progress towards its destination. 
As such, the AV can be classed as a \textit{responder\/} DUV, i.e.\ a DUV that reacts to lower-level stimulus observed on its interfaces with the surrounding environment in order to maintain legally correct driving behaviour and follow the social norms associated with road traffic.
%
This paper investigates the benefits of introducing \textit{agency} into the verification environment in order to address the challenges of verifying the responder DUV. 
Each software agent is tasked with specific goals that aim to achieve verification objectives, e.g.\ reaching coverage targets.
A set of software agents can then be directed to interact, coordinating their behaviour in response to the AV's observed actions in order to increase the likelihood of rare events occurring during simulation to reach coverage targets faster.
%
Our key research question is: what are the benefits of using agent-based test generation for the verification of AVs in simulation?
In particular, we are interested in how agency-directed test generation compares to pseudo-random test generation techniques wrt.\ the following criteria for a 'good' test case, which are inspired by~\cite{fewster1999software}:
%
\textit{effectiveness} in detecting failures, 
\textit{efficiency} in minimising the number if tests required to achieve verification goals, 
\textit{economy} in terms of resource usage 
and also 
\textit{robustness} towards changes. 
Our results suggest that generating tests using directed agents significantly improves upon random and simultaneously provides driving scenarios that are more realistic than those obtained by random test generation.

We regard agent-based test generation as a contribution to the well-established model-based test generation paradigm. A taxonomy of model-based test generation from~\cite{utting2012taxonomy} is given in Figure~\ref{f:taxonomy}.
The agent-based technique creates two new entries in that taxonomy, a new Paradigm under Model Specification, \textit{Agent-based}, and a new Technology under Test Generation, \textit{Agency}, which includes reactive reasoning, causality and strategic planning between multiple agents.
In this paper we use an agent-based model to specify the test environment of the AV. Agency is given to the key dynamic entities in the test environment that interact with the AV, in our case pedestrians. Agency, implemented through a belief-desire-intention agent interpreter such as Jason~\cite{bordini2005jason} is then employed to generate tests based on the multi-agent system that represents the test environment. Note that other test generation techniques, such as random generation, which we use as baseline for evaluation, or model checking~\cite{Bordini2006}, can also be applied to an agent-based model.	

This paper is structured as follows. In the next section we introduce the terminology we will adopt throughout the paper and present the testbench architecture used in our experiment. In Section~\ref{s:background} we review related work on test generation for simulation-based AV verification and introduce the basics of multi-agent systems. Section~\ref{s:case-study} presents our case study, which is centred around test generation for a collision avoidance scenario. Results are presented in Section~\ref{s:results} and evaluated in Section~\ref{s:evaluation}. We conclude in Section~\ref{s:conclusion} and give an outlook on future work. 

\section{Terminology and Testbench Architecture}\label{s:testbench}

We will adopt the terminology defined in~\cite{Ulbrich2015}, where \textit{scene} refers to all static objects including the road network, street furniture, environment conditions and a snapshot of any dynamic elements. Dynamic elements are the elements in a scene whose actions or behaviour may change over time, these are considered actors and may include other road vehicles, cyclists, pedestrians and traffic signals. 
The \textit{scenario} is then defined as a temporal development between several scenes which may be specified by specific goals and values.
A \textit{situation} is defined as the subjective conditions and determinants for behaviour at a particular point in time. 

\begin{figure}[!t]
	\centering
\includegraphics[width=0.98\textwidth]{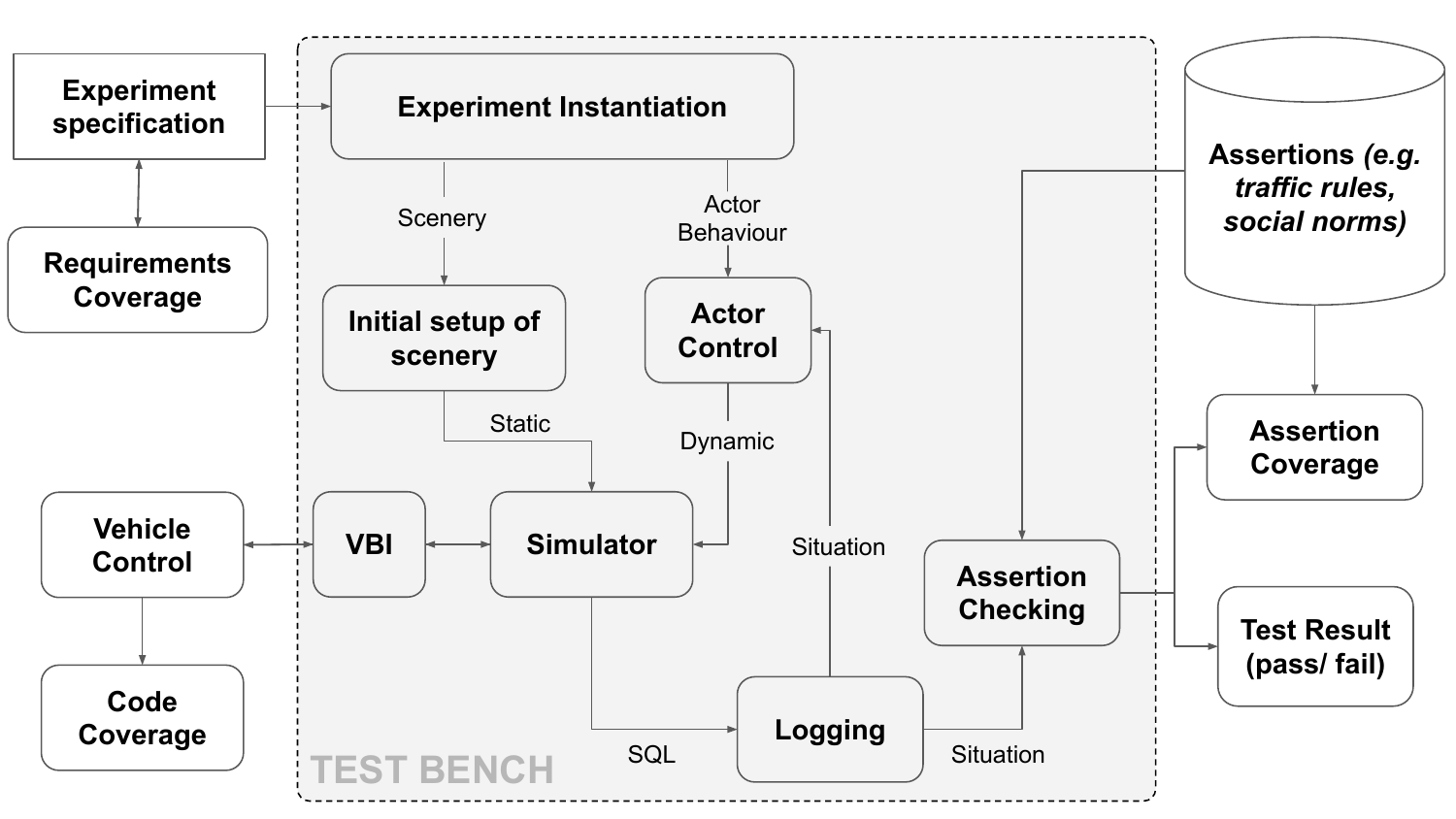}
	\caption{Testbench Architecture.}
	\label{f:testbench}
\end{figure}

The proposed testbench, see Fig.~\ref{f:testbench}, is driven by a specification for the experiment which defines the scene and scenario including all dynamic actors. 
The experiment or test case specification specifies the test inputs, i.e.\ the execution conditions for an item to be tested~\cite{StandardsBoard1990}.
The vehicle behaviour interface (VBI) connects the AV controller (vehicle control) to the simulator. It provides the simulator with the driving decisions of the AV and forwards updates on the scene to the AV controller. 
A geospacial database logs the AV and all other actors to enable post-simulation assertion checking.

Given the experiment specification, tests may be generated in a variety of ways that differ mainly in their effectiveness and efficiency. 
Manually generated tests are typically effective in achieving verification objectives, but are considered expensive due to the high engineering cost involved. 
Random methods are usually employed at the early stages of testing to build up coverage quickly. They suffer, however, from a high number of invalid tests being generated and, even when constrained to produce only valid tests, the tests generated are often not interesting wrt.\ the verification objectives, in our case this refers to exercising the collision avoidance decision making logic of the AV. 
Model-based test generation offers an alternative that produces valid and interesting tests at the cost of developing a model that faithfully encodes the behaviour of the test environment. This model can then be explored in a variety of ways, see Figure~\ref{f:taxonomy}.
%
Our paper explores how the agency that is naturally present in the multi-agent system that represents the dynamic actors in a scene can be used for model-based generation of test scenarios in the context of simulation-based AV verification. 

One could argue that the simulation environment presented to the AV should be as close to the real world it seeks to emulate as the most realistic proving ground. Such efforts seek to reduce the \textit{reality gap}, 
providing the most likely scenarios and actor behaviour. This can be achieved by monitoring real traffic scenarios, e.g.\ tracking individual vehicles, cyclists and pedestrians, and building behavioural models for each of the tracked entities~\cite{behbahani2019learning}.
On the other hand, the creation of edge cases is critical to reach verification objectives in a timely manner. Edge cases are events that occur rarely under normal circumstances, yet realistic \textit{and\/} critical for gaining confidence in the correct behaviour of the AV.
This is exactly what the agent-based test generation approach introduced in this paper is aimed at.

\section{Background}\label{s:background}
\subsection{Related Work}

An overview of the challenges currently faced in software testing of AVs is given by Koopman, highlighting the importance of fault injection into the testing domain~\cite{Koopman2016} and how to decide what aspects of the system need to be tested in the areas of operational design domains, event detection and vehicle manoeuvres~\cite{Koopman2019}. 
Describing a driving scene in natural language has been shown to be an effective framework for scenario generation. This can also be automated based on formalised rules and knowledge~\cite{Bagschik2018} or even evolved from ISO safety standards~\cite{Menzel2018}. 
Hallenbach et al.\ take the approach to test generation of developing composite metrics for traffic quality and using them to determine if a scenario is `critical' or not and therefore worth exploring further~\cite{Hallerbach2018}. 
Generating targeted test cases can be a more efficient way to achieve verification objectives. 
For example, Mullins et al.~\cite{Mullins2018} describe a test generation method that is focused on exploring the transitions between decision making performance modes of the AV, with the aim to find tests that exercise the complete decision making logic. 
Saigol et al.~\cite{Saigol2018} discuss the need for \textit{smart actor control} which they expect may lead to interesting scenarios when these interact with the AV a testbench framework for automated testing. 
Rocklage et al.~\cite{Rocklage2017} approach the problem of test generation from the viewpoint of the other traffic participants, using a trajectory planner to ensure coverage over a fixed list of scene parameters. 
Tuncali et al.,~\cite{Tuncali2019} use rapidly exploring random trees to explore boundary scenarios for different adversarial road users.
Exploiting agency for test generation, however, has not yet been investigated.

\subsection{Multi-Agent Systems}

Georgeff and Lansky were key in the development of the belief-desire intention (BDI) agent programming approach~\cite{georgeff1987reactive} and also in the early work of multi-agent systems~\cite{georgeff1988communication}.
In a multi-agent system, multiple intelligent agents interact in order to collectively solve problems that are too difficult or impossible for individual agents to achieve. 
A multi-agent system includes a set of software agents and their environment. Agents can be equipped with varying degrees of agency, starting from  passive agents, such as obstacles, including active agents that have goals of low complexity, such as pedestrians that act as part of a group, to rational agents that are capable of reasoning based on a cognitive model of the situation, their perception of the surrounding environment, and a set of rules they can use for strategic planning and communication with other agents in the system.
The individual agents are considered \textit{autonomous} and in control of their behaviour within the multi-agent system as they pursue their goals. 
This results in self-organisation and self-direction towards achieving a common goal. 

Combining the BDI framework with an automated test generation approach leads to the concept of a software agent capable of generating tests.
Such intelligent, agent-based test generation has first been applied in the human-robot interaction domain~\cite{Araiza-Illan2016}, where a coverage-driven test generation approach was supplemented with reinforcement learning to improve the efficiency and effectiveness of testing the critical part of a robot-to-human handover task within a collaborative manufacturing scenario. 
Test agents have also been proposed by Enoiu et al.~\cite{Enoiu2019} for regression testing, although they are used more for test selection from a library of tests, rather than for test generation. These agents use inter-agent messaging to decide what tests to execute and with what prioritisation.

\section{Case Study} \label{s:case-study}
 
This section describes a case study to explore the proposed agent-based test generation method. \footnote{The code used is available at \url{github.com/TSL-UOB/CAV-MAS}.} Our aim is to generate tests that exercise an assertion that requires the AV to avoid collisions with other road users, provided that such road user, a pedestrian in this case, intrudes in the path of the AV in such a way that a collision can be avoided by the AV either by braking or manoeuvring. This is similar to standardised testing, e.g. Euro-NCAP CPNA-25~\cite{EURO-NCAP}. In Fig.~\ref{gridRoad} this is illustrated by the \textit{precondition zone}, i.e.\ the area of interest in which the assertion is activated.

In the following, we shall term tests that activate the assertion \textit{successful tests}; these count towards assertion coverage. 
Intrusions that fall within the \textit{stopping distance} of the AV~\cite{codes2015highway}, i.e.\ the 12m long zone directly ahead of the AV when driving at 30km/h, equivalent to {\raise.17ex\hbox{$\scriptstyle\sim$}}20mph (miles per hour) as marked in Fig.~\ref{gridRoad}, are not considered interesting as they result in unavoidable collisions; tests of this type are of limited use~\cite{Tuncali2018}. Thus, in this case study, unavoidable collisions are those with a time to collision (TTC) of less than 1.33s. The extent of the \textit{precondition zone} is limited to a single simulation tick (1.0s) which puts the upper limit at 2.33s ahead of the \textit{stopping distance}.

A single assertion is chosen so that we can study the fundamental properties of agent-based test generation in a simple setting in comparison to using random test generation techniques. Moreover, in this investigation we do not consider the test result (pass or fail), i.e.\ the behaviour of the AV in response to the assertion being activated.

\subsection{Test Environment}
\begin{figure}[!t]
	\centering
\includegraphics[width=0.98\textwidth]{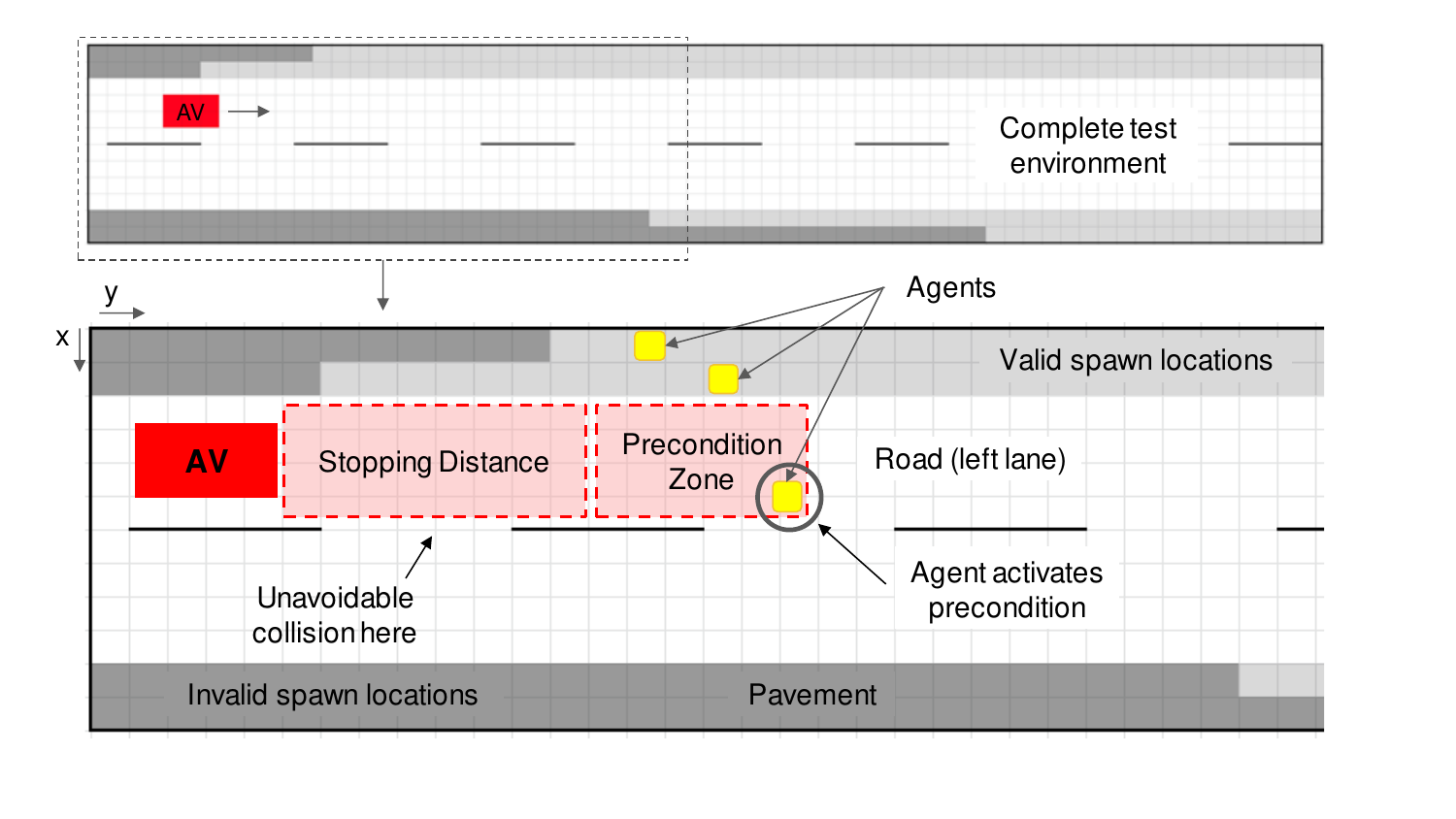}
	\caption{Test environment at full scale (upper part) and in detail (lower part), marking valid and invalid spawn locations for the pedestrians and indicating the position and direction of the AV (upper part).}
	\label{gridRoad}
\end{figure}

The test environment, depicted in Fig.~\ref{gridRoad}, is a straight two-lane 99m road with two 6m wide lanes. Pavements are on both sides of the road which are three meters in width giving a total area of 18m$\times$99m. The AV velocity is 9.0m/s (equivalent to {\raise.17ex\hbox{$\scriptstyle\sim$}}20mph) and the pedestrian velocity is 1.4 m/s (equivalent to {\raise.17ex\hbox{$\scriptstyle\sim$}}3mph). The map is discretised with 1.5m resolution for simple division into the AV and pedestrians' velocities. Thus, in the discretised world the AV velocity is 6 cells/s and the pedestrians' velocity is roughly 1 cell/s. The total number of map cells is 12$\times$66 = 792 cells. 

The AV travels along the left-hand lane of the road starting at cell $y=0$ and travelling to the end, cell $y=66$. If the assertion is triggered
or the AV reaches the end of the road, then the test is restarted. The AV occupies an area 4.5m$\times$3m equivalent to 3$\times$2 cells. There are no other vehicles and the right-hand side of the road is unoccupied.

To activate the assertion, the precondition must be satisfied, i.e.\ an agent has entered the \textit{precondition zone} of the AV, which extends 6 cells forwards of the \textit{stopping distance} of the vehicle as shown in~Fig.~\ref{gridRoad}. Under the given conditions, it is impossible for an agent to move into the AV's path from some pavement areas within the test environment. These are marked in dark grey in Fig.~\ref{gridRoad} and labelled invalid spawn locations. 

The environment is initialised with the agents spawned randomly on any valid spawn location within the pavement area. When the environment is reset, new random locations are chosen for the agents. Random locations are controlled by using a fixed seed that is based on the experiment number; thus, initial spawn locations become repeatable (through using the same seed), which ensures a fair comparison between different experiments.

\subsection{Test Agents}

The test agents in our case study are pedestrians. At the start of a test, pedestrians are randomly located onto the pavement area using valid spawn locations. The AV is a vehicle that drives at constant speed in a straight line, it does not brake or turn. For the purpose of comparing the performance of agent-based test generation techniques, the test agents have different behaviours from simple random (i.e.\ no direction through agency) to more directed, strategic planning-based agency, details of these are provided in this section.

We distinguish two main classes of behaviours, random and agency-directed, see also Table~\ref{t:ResultsTable}. For the random class, \textit{random} behaviour means that the test agent can perform any random action at each simulation tick, with the available actions being: do nothing (stand still), move forward, backward, left or right. In \textit{constrained random} mode, the pedestrian is initially walking along the pavement and has only one action available; to randomly choose when to cross the road. The constrained random behaviour is included so that a comparison between crossing the road at a targeted or at a random time can be made. 

The second class of behaviours is agency-directed, taking into account the agent's perceptions for strategic planning. Agents are initially walking along the pavement as in constrained random mode. The \textit{proximity} behaviour instructs the agent to cross the road when the AV is within a certain radius. The \textit{election} behaviour ranks agents within a proximity radius to the AV and elects the agent with the shortest distance to the AV. The major difference between the agent-directed behaviours is that the \textit{election} behaviour will only elect a single agent to cross the road whereas the \textit{proximity} behaviour will result any number of agents crossing the road as long as they are within range. The trigger radius is 15 cells using city-block measure and is the same value for both \textit{proximity}  and \textit{election} behaviours. While the \textit{proximity} behaviour may lead to the desired coverage it is less realistic than the \textit{election} behaviour.

The final factor to consider is the number of agents per test; too few and the likelihood of an agent activating the assertion will be low and entirely dependent on the initial starting position of the agent in the test environment due to differences in speed. The number of agents is physically limited by the number of available grid locations (1.5m spacing) on the pavement. As the number of agents, $nA$, increases, it is expected that the probability of activating the assertion using a random behaviour increases significantly as more agents appear in the road, with a corresponding increase in computational expense. In comparison, when using more intelligent, agency-directed behaviour, we expect that with increasing $nA$, agents that can activate the assertion precondition are being generated more readily, thus resulting in shorter, less complex and hence more efficient tests. The range of $nA$ explored in this case study is from 1 to 20.

\subsection{Scoring}

In an attempt to encourage more realistic agent behaviour, a basic scoring system is used to direct agent behaviour by penalising or rewarding certain actions. In particular, a living cost of one is charged for each elapsed time step to promote shorter tests, and a penalty of five is issued for each time step that an agent spends in the road. This is based on the general observation that most pedestrians do not predominately walk in the road but rather cross over it and hence higher scores are associated with less time spent in the road. A reward of 100 is given if the agent enters the AV's \textit{precondition zone} More sophisticated scoring systems will be explored in our future work. 

\subsection{Simulation and Logging}

Each agent behaviour was implemented in Jason~\cite{bordini2005jason} and executed within the testbench environment as described in Section~\ref{s:testbench}.
During simulation, the agents' actions, scores and time to test completion were recorded in a log. Each agent behaviour was repeated 1000 times and the number of successful tests was counted. A successful test is one that has a pedestrian intrude into the \textit{precondition zone} of the AV, thereby creating the opportunity for a subsequent intersection with the AV, i.e.\ a collision, thus triggering the AV's collision avoidance decision making logic. A list of random start locations was created for the pedestrians for all settings of agent numbers, $nA$. This list was used to spawn the pedestrians for each of the agent behaviours to ensure the initial conditions between experiments were identical. 

\section{Results}\label{s:results}
The results are evaluated based on five criteria; Accuracy is the ratio of successful tests the agents generated to the number of all tests, Score is a measure of how natural the agents behaved, Combined Score combines accuracy with score, Agent CPU Time is a measure of computational efficiency, and Test Generation Time is how long the agents took to generate the test in both simulation ticks and wall clock time. Both Score and Test Generation Time have distributions associated with them and therefore confidence intervals are provided.

\subsection{Test Accuracy}
\begin{figure}[!t]
	\centering
\includegraphics[width=0.98\textwidth]{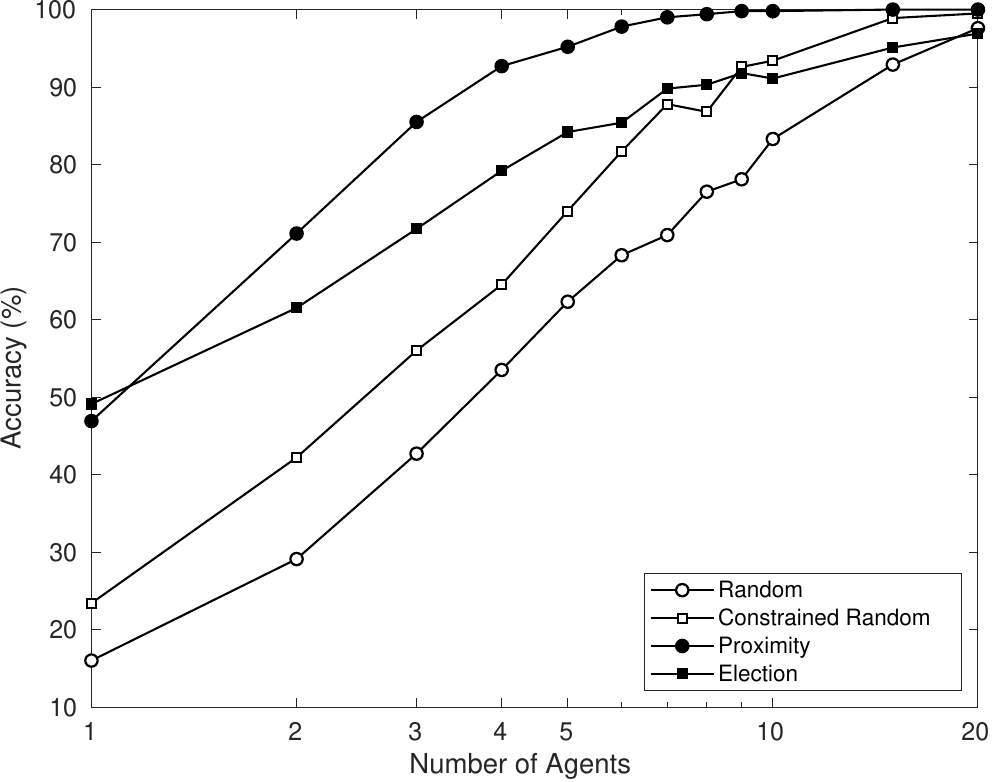}
	\caption{Accuracy of agent behaviours to generate successful tests.}
	\label{f:accuracy}
\end{figure}

Test accuracy, defined as the number of tests that have activated the precondition for the assertion as a ratio of all tests, is shown for each agent behaviour in Fig.~\ref{f:accuracy}. The \textit{random} and \textit{constrained random} behaviours have significantly lower accuracy compared to the directed agent behaviours when $nA<10$. For high $nA$ the accuracies converge mostly due to a saturation of agents in the test environment. Fig.~\ref{f:accuracy} also shows that for $nA=1$ a directed agent outperforms a random agent by more than 2:1. The \textit{election} behaviour generates a slightly higher accuracy than the \textit{proximity} agent behaviour, but only by 2.2\%. However, for $nA=2$ and above the \textit{proximity} agent behaviour shows a 10\% increase in accuracy over the \textit{election} behaviour; this is because agents with this behaviour have multiple attempts to trigger the precondition whereas the election behaviour permits only one. 

\subsection{Test Score} \label{s:testscore}

\begin{figure}[!t]
	\centering
\includegraphics[width=0.98\textwidth]{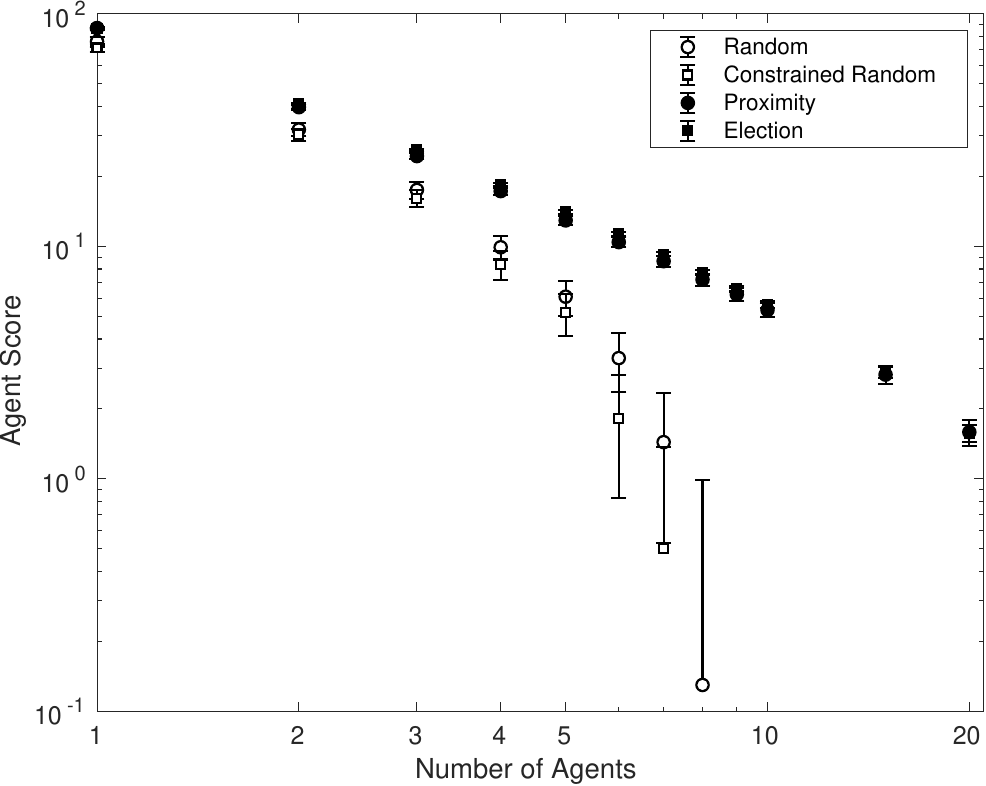}
	\caption{The average agent score for successful tests with 95\% confidence intervals for each agent behaviour.}
	\label{f:agentscore}
\end{figure}

For tests that activate the precondition the average agent score is compiled including 95\% confidence intervals, see Fig.~\ref{f:agentscore}. The maximum theoretical score of any agent is 94, which includes 100 points for a successful test subtracting a living cost of one and a road penalty of five; this requires the agent to spawn adjacent to the \textit{precondition zone}. For a single agent, scores are similar for all agent behaviours, as only successful tests are included in the scoring. As the number of agents increases, the random behaviours diverge from the directed behaviours, and the variance observed for the random behaviour scores increases. This indicates that the directed agents display more natural behaviour than those using random.

As $nA$ increases, the random agents are more often found in the road and hence their average score decreases. In contrast, the directed agents are only found crossing the road when they deem fit, and they remain on the pavement at all other times, keeping the score much higher and resulting in more realistic test cases. The basis for this rationale is from the general observation that people walk on pavements rather than roads and hence walking in the road is penalised. No significant difference between the two directed behaviours can be observed in the score, even with a high number of agents. This indicates that  both result in similar level of realism in agent behaviour, at least within the limited scope of this example.

\subsection{Combined Score}
\begin{figure}[!t]
	\centering
\includegraphics[width=0.98\textwidth]{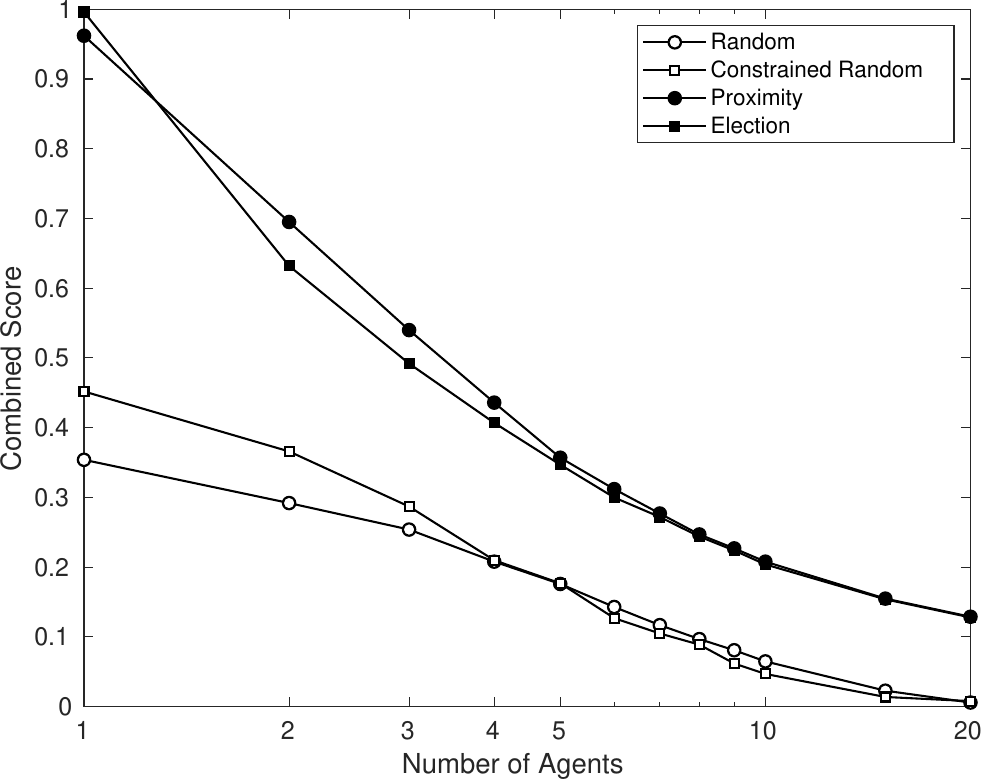}
	\caption{The combined score including accuracy and the original score for each agent behaviour.}
	\label{f:combined}
\end{figure}

As the score in Section~\ref{s:testscore} only includes tests that meet the precondition, it is tempting to interpret these results as overly optimistic. To counteract this we use a \textit{combined score} that is calculated based on the score and accuracy using (score * accuracy / 1000). This combined measure promotes scores that are attached to high accuracies, describing agents that can generate successful tests with natural pedestrian behaviour. The normalised results, Fig.~\ref{f:combined}, show that the directed agents are more than twice as effective within this new combined definition than agents with random behaviour for $nA=1$, although this lead drops rapidly with increasing agent numbers, reaching a steady performance gap of around 12\%. The \textit{constrained random} agent behaviour outperforms the \textit{random} for $nA<4$ beyond which there is little difference. Therefore, if random behaviour is chosen, then the extra effort in implementing constrained random may not be worthwhile considering the low returns, except when using low agent numbers. The \textit{election} agent behaviour has the highest combined score for $nA=1$, but as agent numbers increase, better results can be seen for the \textit{proximity} agent behaviour up to $nA=4$, beyond which there is no significant difference between the two directed agent behaviours.

\subsection{Agent CPU Time}
\begin{figure}[!t]
	\centering
\includegraphics[width=0.98\textwidth]{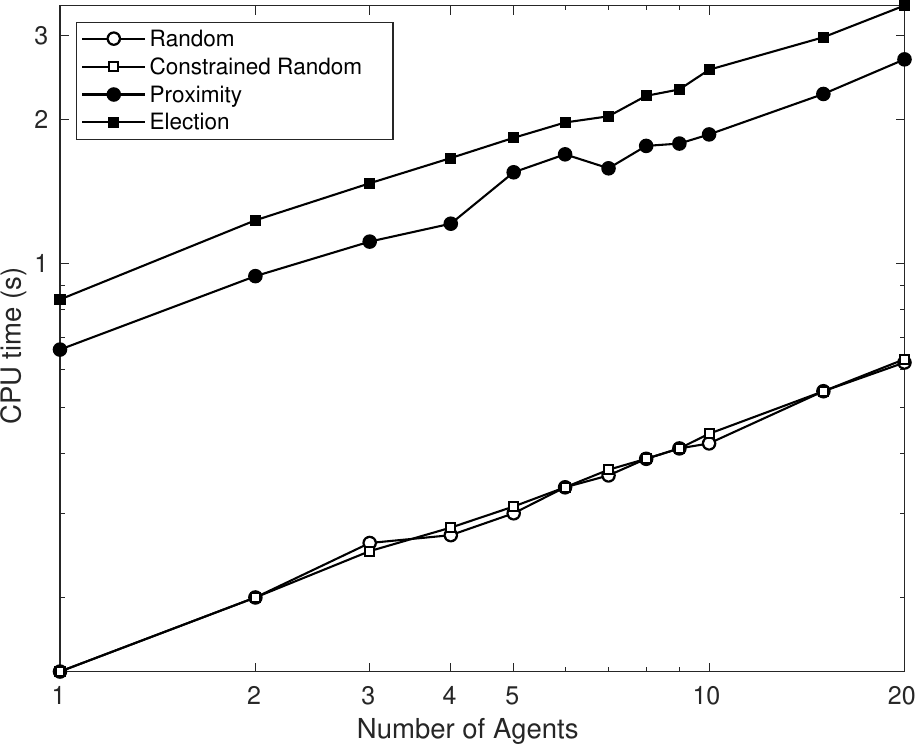}
	\caption{The CPU time, $t_{C}$, to execute agent actions averaged over 1000 runs.}
	\label{f:cputime}
\end{figure}

For each successful test, the CPU time used to execute the agent actions was averaged over the 1000 runs and assessed across different agent behaviours and numbers of agents, Fig.~\ref{f:cputime}. This allows comparing the resources required to execute agents of different behaviour types. The more complex agents have additional CPU overhead compared to agents with random behaviour. This difference remains relatively stable as agent numbers increase.

\begin{table*}
\centering
\caption{Test agent summary table showing description and number of lines of code ($LOC$) for each agent and sample of results: Accuracy (ACC), Combined Score ($s_c$), CPU Time ($t_{c}$) and Test Generation Time ($t_{g}$) for $nA=3$.}
\label{t:ResultsTable}
\begin{tabular}{|p{2.1cm}|p{4.0cm}|r||c|c|c|c|}
\hline
\textbf{Behaviour} & \textbf{Agent Description} & $LOC$ & ACC (\%) & $s_c$ (points)&  $t_{c}$ (s) & $t_{g}$ (ticks) \\
\hline
Random & Randomly perform action (forward, backward, left, right, stop) &  23& 42.7 & 0.254 & 0.09 & 9.11 \\
Constrained Random & Walk along pavement, randomly cross the road 		&  82& 56.0 & 0.287 & 0.08 & 8.83 \\
Proximity & Walk along pavement, cross road when AV in range 			&  86& 85.5 & 0.540 & 0.37 & 6.79 \\
Election & As in Proximity but elect a single agent to cross 			& 235& 71.7 & 1.470 & 0.49 & 6.59 \\
\hline 
\end{tabular}
\end{table*}

\subsection{Test Generation Time}
\begin{figure}[!t]
	\centering
\includegraphics[width=0.98\textwidth]{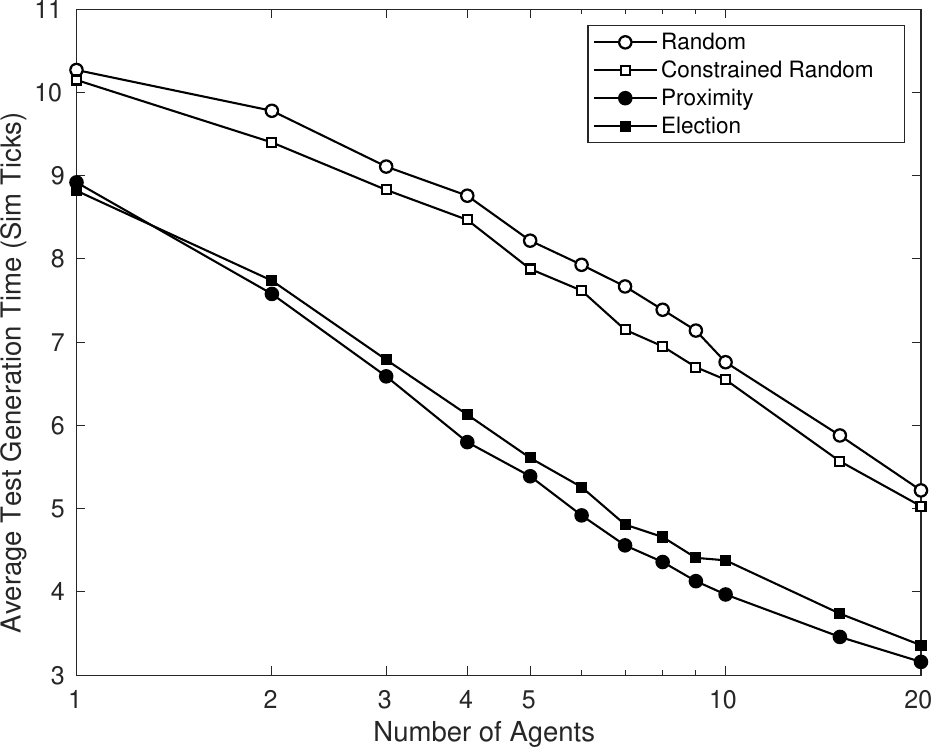}
	\caption{The average time taken for an agent to find a successful test, $t_{g}$.}
	\label{f:time}
\end{figure}

For each successful test generated, the number of simulation ticks consumed during test generation was averaged over 1000 runs and compared across different agent numbers for all four agent behaviours, Fig.~\ref{f:time}. 
The directed agents show improved performance over the agents with random behaviour for $nA=1$ by approximately a single simulation tick and this trend continues as agent numbers increase. By $nA=20$ the directed agents are 1.85 simulation ticks faster than the agents with random behaviour. Overall, the \textit{proximity} agents find tests in the smallest number of simulation steps.

\section{Evaluation}\label{s:evaluation}
The results are assessed against the four criteria identified in Section~\ref{s:introduction}.

\textit{Effectiveness and Efficiency}: How effective is the method at generating tests that detect failures in the DUV, and how efficient is the method in minimising the number of tests required to achieve verification goals (i.e.\ assertion coverage)? The accuracy metric, defined above, shows how often an agent generates a test that satisfies the precondition of the assertion, thereby triggering the execution of the AV's collision avoidance decision-making logic, creating the potential to reveal defects in the DUV and achieving assertion coverage in the process. Fig.~\ref{f:accuracy} shows that a small number of directed agents is around twice as effective as random ones and over three times as effective as a single agent.

\textit{Economy}: How costly is the test case to develop and run? 
The resource cost (CPU time) to execute the agent actions, Fig.~\ref{f:cputime}, is 4-5 times higher for the directed agents than for agents with random behaviour, see also heading $t_{c}$ in Table~\ref{t:ResultsTable} for $nA=3$. However, the test generation time ($t_g$), Fig.~\ref{f:time}, indicates that on average the directed agents find successful tests faster than random. Therefore, although more resource needs to be  allocated to the directed agents, overall they find test cases faster due to higher efficiency. Wrt.\ development effort, the lines of code for each agent behaviour, see Table~\ref{t:ResultsTable}, show that for a moderate investment the \textit{random} behaviour can be improved significantly to obtain the performance of the \textit{proximity} behaviour. However, diminishing returns are evident beyond that. This simple scenario would suggest that the level of agent complexity should be considered carefully as a simpler level of agency could potentially be more beneficial than more complex options. 

\textit{Robustness} refers to the maintenance required to adapt tests to software changes, i.e.\ different \textit{scenes}. In the case study each test generated is of a different scene as the agent positions are randomly generated. The directed agents show higher robustness based on their accuracy when compared to random behaviours. 

Overall, our results show that agency-directed testing outperforms random based, confirming that even a small amount of agency can be a distinct advantage over random techniques. 

\section{Conclusion and Future Work}\label{s:conclusion}
The MAS programming paradigm offers rational agency, causality and strategic planning to software agents; these have been exploited in this research for test generation. On the example of a key assertion on collision avoidance we show that, by encoding a variety of different behaviours respondent to the agent's perceptions of the test environment, the agency-directed approach generates twice as many effective tests than a pseudo-random approach. Furthermore, agents can be encoded to behave more realistically without compromising their effectiveness. Our results suggest that generating tests using testing agents is a promising avenue of research and has been shown to significantly improves upon random whilst simultaneously providing more realistic driving scenarios.

Future work will extend this initial study to multiple assertions and a significantly larger variety of scenes including different road networks. 
As discussed in~\cite{Eder2007}, agents could have their goal selection modified based on coverage feedback, giving rise to  agent-based coverage-directed test generation. Abstracting agent perceptions to a feature based representation could ensure the agent state space scales up to large physical maps and is adaptable to new features as more assertions are added. Including personality in agents is also another avenue that could provide a tuning parameter~\cite{Zoumpoulaki2010} to generate edge cases. 

\section{Acknowledgement}
This research has in part been funded by the ROBOPILOT and CAPRI projects. Both
projects are part-funded by the Centre for Connected and Autonomous
Vehicles (CCAV), delivered in partnership with Innovate UK under grant numbers
103703 (CAPRI) and 103288 (ROBOPILOT), respectively.

\bibliography{ait-2020}

\end{document}